\documentclass[twocolumn,prb,showpacs,amsmath,amssymb]{revtex4}

\usepackage{graphicx}
\usepackage{bm}
\usepackage{epsf}

\newcommand{\beq}{\begin{equation}}
\newcommand{\eeq}{\end{equation}}
\newcommand{\beqa}{\begin{eqnarray}}
\newcommand{\eeqa}{\end{eqnarray}}
\newcommand{\beqan}{\begin{eqnarray*}}
\newcommand{\eeqan}{\end{eqnarray*}}
\newcommand{\ben}{\begin{enumerate}}
\newcommand{\een}{\end{enumerate}}
\newcommand{\bit}{\begin{itemize}}
\newcommand{\eit}{\end{itemize}}

\newcommand{\ep}{\thinspace . }

\newcommand{\mathbfx}[1]{{\mbox{\boldmath $#1$}}}

\newcommand{\x}{\mathbfx{x}}

\begin{document}


\title{A finite element analysis of a silicon based double quantum
dot structure}
\author{S. Rahman$^1$}
\email{S.Rahman.00@cantab.net}
\author{J. Gorman$^1$}%
\author{C. H. W. Barnes$^1$}%
\author{D. A. Williams$^2$}%
\author{H. P. Langtangen$^3$}%
\affiliation{$^1$Cavendish Laboratory, Cambridge
University, J J Thomson Avenue, Cambridge, CB3 OHE, UK}%
\affiliation{$^2$Hitachi Cambridge Laboratory, J J Thomson Avenue, Cambridge, CB3 OHE, UK}%
\affiliation{$^3$Simula Research Laboratory, Martin Linges v 17,
Fornebu P.O.Box 134, 1325 Lysaker, Norway}%

\date{\today}

\begin{abstract}

We present finite-element solutions of the Laplace equation for
the silicon-based trench-isolated double quantum-dot and the
capacitively-coupled single-electron transistor device
architecture. This system is a candidate for charge and spin-based
quantum computation in the solid state, as demonstrated by recent
coherent-charge oscillation experiments. Our key findings
demonstrate control of the electric potential and electric field
in the vicinity of the double quantum-dot by the electric
potential applied to the in-plane gates. This constitutes a useful
theoretical analysis of the silicon-based architecture for quantum
information processing applications.

\end{abstract}


\maketitle

Recent experiments conducted on trench-isolated double quantum-dot
(IDQD) structures have successfully demonstrated detection of
single-electron polarization,\cite{Emiroglu:03} and
coherent-charge oscillation.\cite{Gorman:05} This highlights the
possibility of constructing charge-based quantum computer circuits
in Si, with coherence times of the order
$100$~ns.\cite{NielsonBook,Loss:98} The architecture for a single
qubit device is a complex, three-dimensional structure consisting
of a single-electron transistor (SET), an IDQD, and gate
electrodes. This makes it difficult to determine theoretically the
system evolution by means of a complete and self-consistent
Schr\"{o}dinger-Poisson analysis. This is particularly the case if
the analysis were to fully take into account the device geometry
and all interactions while performing quantum manipulations within
the coherence time of the qubit.\cite{ft1}

In this paper, we present a significant contribution to such an
analysis by a finite-element solution of the Laplace equation with
the three-dimensional device geometry and material composition
taken into account. The aim of this work is to demonstrate the
electrostatic effect on the IDQD structure when voltages are
applied to the in-plane control gates of the device.

\begin{figure}[!h]
\begin{center}
\rotatebox{0}{\scalebox{0.65}[0.65]{\includegraphics*[0.0in,4.25in][6.0in,8.25in]{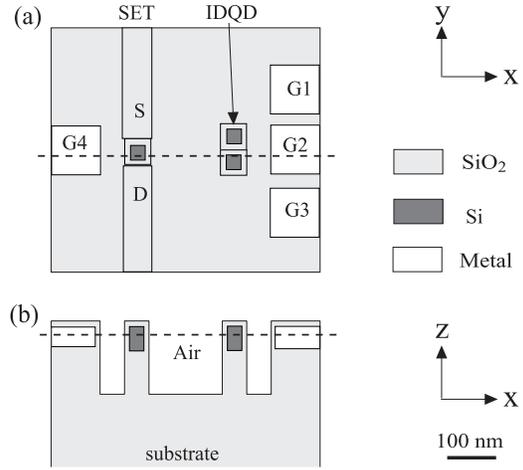}}}
\end{center}
\caption{A schematic representation of the Si IDQD device used in
the numerical simulations. Rectangular approximations are made to
the device elements. (a) A view of the $x$-$y$ plane at the level
of the dashed line in (b). (b) A view of the $x$-$z$ plane at the
level of the dashed line in (a). The trench-like structures are
connected to the in-plane metal gates through a column of air.
Each trench extends about $150~$nm above the oxide base.}
\label{fig:schematic}
\end{figure}

Figures~\ref{fig:schematic}(a) and \ref{fig:schematic}(b) show
device schematics in the $x$-$y$ and $x$-$z$-planes, respectively.
The trench isolation, illustrated in Fig.~\ref{fig:schematic}(b),
is formed by high-resolution electron-beam lithography and
reactive-ion etching. Each trench is approximately $150$~nm deep
and runs into the buried-oxide (BOX) layer of the
silicon-on-insulator (SOI) wafer. The active regions of the device
elements are P doped Si, which are electrically isolated from
other device elements, as seen in Fig.~\ref{fig:schematic}. In
this work, we use rectangular approximations to the device
elements and the etched profiles to simplify the
analysis.\cite{ft2}

The small dimensions of the quantum dots and the $20$~nm
constriction between them, which is fully depleted and acts as a
tunable inter-dot tunnel barrier, result in a significant
double-well type confinement potential for the electrons that
occupy discrete quantum states on each quantum
dot.\cite{Fujisawa:98,vanDerWiel:02} The tunable inter-dot
coupling causes the wave functions in the two dots to overlap and
hybridize so that they may be thought of as pseudo-molecular
states of an artificial two atom molecule. The IDQD is an
electrically-isolated component that is coupled only capacitively
to the rest of the circuit, including the SET for read-out, and
the in-plane control gates (G1 to G3) for manipulation. Voltages
applied to the gates G1 to G3 are used to tune the electric field
in the vicinity of the IDQD, and thus, the confinement potential
asymmetry and inter-dot tunnelling. Hence, an electron initially
localized on one quantum dot may be allowed to tunnel to the
opposite quantum dot by such manipulation.

An electric field is induced on the SET as a result of this
polarization process. This modulates the chemical potential of the
SET, and, therefore, the conductance through the source and drain
leads under a finite SET bias condition. To maximize the change in
conductance, the SET is initially tuned to the charge-sensitive
regime by an appropriate voltage bias at G4, and under a small
source-drain bias to ensure operation in the linear transport
regime. Such manipulation of the device has only been shown
through experiment so far. Therefore, a thorough theoretical
analysis is necessary to complement the recent experimental
findings, and build a more comprehensive understanding of the
physical mechanisms involved.

Analytic methods exist for calculating the electrostatic potential
in two-dimensional electron gases generated by patterned surface
gates on GaAs/AlGaAs heterostructures.\cite{Davies:94,Davies:95}
While these analytic methods yield useful results for such
devices, they are unsuitable for trench-isolated Si structures,
where the geometry is much more sophisticated. Therefore,
numerical methods have to be implemented. The finite-element
method is a well-suited means for simulation of
geometrically-complicated domains,\cite{Cook,Zienkiewicz} and is
commonly used to solve Poisson-type
equations.\cite{Iserles,DpBook2}

In order to determine the electric field throughout the modelled
device regions, the numerical solution to the Laplace equation in
three-dimensions is performed:

\begin{eqnarray}\label{eqn:poisson}
     \nabla \cdot [ \epsilon (\mathbf{x}) \nabla \phi (\mathbf{x}) ] &=& 0, \qquad \mathbf{x} \in \Omega \in \mathbb{R}^3, \label{eqn:lap1} \\
    \phi(\mathbf{x}) &=& D_i, \qquad \mathbf{x} \in \partial \Omega_{D_i}, \label{eqn:lap2}\\
     - \epsilon (\mathbf{x}) \frac{\partial \phi }{\partial n} &=&
     q,\qquad \mathbf{x} \in \partial \Omega_{N}, \label{eqn:l3}
\end{eqnarray}

\noindent where $\phi (\mathbf{x})$ is the electrostatic potential
and $\epsilon (\mathbf{x})$ is the material dielectric parameter.
The dielectric parameter varies discontinuously on moving through
the different materials - air ($\epsilon_0 = 8.85 \cdot 10^{-12}
~$Fm$^{-1}$), Si ($11.0 \epsilon_0$) and SiO$_2$ ($4.5 \epsilon_0
$). We apply Dirichlet boundary conditions to the surfaces of the
metal gates and to the grounded base of the device, and the
Neumann boundary condition to the exposed surfaces. (We apply
Neumann boundary conditions with $q=0$, but for generality, we
include in our discussion the possibility of non-zero $q$).

The finite element solution of the Laplace equation is well
covered in the literature.\cite{Iserles,DpBook2,Mohan} The basic
idea of the finite element method is to approximate the unknown
fields, for example $\phi$ in the Laplace equation above, by
$\widetilde{\phi}$ which is a combination of linearly independent
basis functions $N_j$

\begin{equation}\label{eqn:fem4}
\widetilde{\phi}(\x) = \sum_{j=1}^M \phi_j N_j(\x) ,
\end{equation}

\noindent where $M$ is the number of basis functions and $\phi_j$
are the expansion coefficients to be determined. In the finite
element method, the computational domain $\Omega$ is divided into
a number of elements, and the $N_j$ are chosen to be piecewise
polynomials such that they are non-zero only in a `few' adjacent
elements.\cite{ft3,Iserles} The method then requires the
substitution of $\widetilde{\phi}$ into the Laplace equation, and
the residual $R = \nabla \cdot [\epsilon \nabla \widetilde{\phi}
]$ to be orthogonal to the space spanned by a linearly independent
set $\{ W_1,\ldots,W_n\}$. In our calculations, we implement the
orthogonality through the weighted residual statement:

\begin{eqnarray}
\int_{\Omega} R W_i d\Omega = 0 , \qquad i=1,\hdots,M ,
\end{eqnarray}

\noindent and Galerkin's method i.e. set $W_i$ equal to the basis
functions $N_i$ to obtain

\begin{eqnarray}\label{eqn:fem1}
    \int_{\Omega} N_i [
    \frac{\partial}{\partial x} \epsilon \frac{\partial \widetilde{\phi}}{\partial
    x}
    +
    \frac{\partial}{\partial y} \epsilon \frac{\partial \widetilde{\phi}}{\partial
    y}
    +
    \frac{\partial}{\partial z} \epsilon \frac{\partial \widetilde{\phi}}{\partial
    z}]
    d\Omega  = 0 \ep
\end{eqnarray}



Using integration by parts, we reduce the order of derivatives in
Eq.~(\ref{eqn:fem1});

\begin{eqnarray}\label{fem2}
   -  \int_{\Omega}
     \epsilon \frac{\partial \widetilde{\phi}}{\partial x} \frac{\partial N_i}{\partial x}
    +
     \epsilon \frac{\partial \widetilde{\phi}}{\partial y} \frac{\partial N_i}{\partial y}
    +
    \epsilon \frac{\partial \widetilde{\phi}}{\partial z} \frac{\partial N_i}{\partial z}
    d\Omega
    \nonumber \\ \nonumber\\
    +\int_{\partial \Omega} N_i \epsilon \frac{\partial \widetilde{\phi}}{ \partial n} d\Gamma
    =0 \ep
\end{eqnarray}

The weighted residual method and integration by parts leads to a
natural mechanism for the incorporation of derivative boundary
condition given by Eq.~(\ref{eqn:l3}) for the Laplace operator
$\nabla \cdot [\epsilon \nabla \widetilde{\phi}]$. Hence, we
obtain after expanding the approximation for $\widetilde{\phi}$

\begin{eqnarray}\label{eqn:fem5}
      \int_{\Omega}
     \epsilon \frac{\partial N_i}{\partial x} \frac{\partial N_j}{\partial x} \phi_j
    +
     \epsilon \frac{\partial N_i}{\partial y} \frac{\partial N_j}{\partial y} \phi_j
    +
     \epsilon \frac{\partial N_i}{\partial z} \frac{\partial N_j}{\partial z} \phi_j
     d\Omega
     \nonumber\\ \nonumber\\
    +  \int_{\partial \Omega} N_i q d\Gamma
    =0 ,
\end{eqnarray}

\noindent where summation is implied over repeated indices. The
problem has now been reduced to one of matrix inversion;

\begin{equation}\label{eqn:femMain}
    K \Phi = F ,
\end{equation}

\noindent where the `stiffness' matrix $K$ is given by

\begin{equation}\label{eqn:femK}
    K_{ij} = \int_{\Omega}
     \epsilon \frac{\partial N_i}{\partial x} \frac{\partial N_j}{\partial x}
    +
     \epsilon \frac{\partial N_i}{\partial y} \frac{\partial N_j}{\partial y}
    +
     \epsilon \frac{\partial N_i}{\partial z} \frac{\partial N_j}{\partial z}
     d\Omega ,
\end{equation}

\noindent and the RHS vector is given by

\begin{equation}\label{eqn:femF}
    F_i=-\int_{\partial \Omega} N_i q d\Gamma ,
\end{equation}

\noindent and $\Phi$ is simply the vector of unknowns $\phi_j$.
Dirichlet boundary conditions are implemented by forcing
prescribed values of $\phi_j$.


In the formulation and solution of Eq.~(\ref{eqn:femMain}) we have
chosen linear basis functions corresponding to 8-noded brick
elements. This method therefore has a convergence rate for error
of 2.0.\cite{ft4}

We have employed Gaussian quadrature in three-dimensions for the
volume integrals and two-dimensions for surface integrals, and a
conjugate gradient method with Incomplete Lower and Upper (ILU)
factorization preconditioning to solve the resulting linear system
of equations. Sufficient resolution was obtained by a mesh with
$107\times 77\times 9$ nodes in the $x$, $y$ and $z$ directions,
respectively. We implemented adaptive mesh refinement along $z$
axis to improve accuracy in the vicinity of the active region.

\begin{figure}[!h]
\begin{center}
\rotatebox{0}{\scalebox{0.65}[0.65]{\includegraphics*[1.5in,0.1in][6.0in,6.2in]{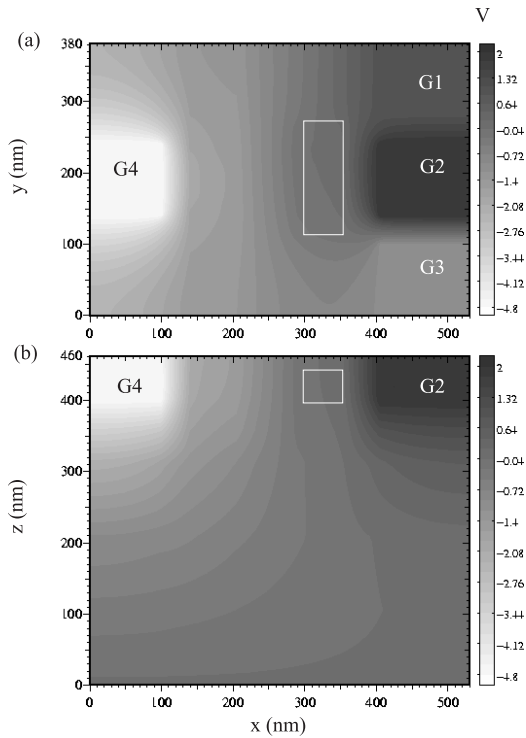}}}
\end{center}
\caption{Two-dimensional slices taken from the full
three-dimensional solution of the Laplace equation. (a) A slice
from the $x$-$y$ plane at $z=420$~nm, through the center of the
gates and IDQD. (b) A slice from the $x$-$z$ plane at $y=190$~nm,
through the center of gate G2. The gate G4, is set to $-4.8$~V.
The gates G1, G2 and G3 are set to $1$~V, $2$~V and $-1$~V
respectively. The position of the IDQD is outlined.}
\label{fig:2dsolutions}
\end{figure}

Figures~\ref{fig:2dsolutions}(a) and \ref{fig:2dsolutions}(b) show
cross-sectional slices along orthogonal planes of the full
three-dimensional solution for the simulated electric potential.
The simulation was performed with the following parameters: the
gate potentials of G1, G2 and G3 are set to +$1$~V, $2$~V and
-$1$~V, respectively; G4 is set to $-4.8$~V. The voltage chosen
for G4 is approximately equal to that used for this gate in the
experimental demonstrations in Ref.~\onlinecite{Emiroglu:03} where
single electron polarization of the IDQD was obtained for such a
device. The choices of G1, G2 and G3, are also similar to those in
the experiments but for this simulation, the exact values are
chosen so that three-dimensional illustrations are as clear as
possible.

Figures~\ref{fig:2dsolutions}(a) and \ref{fig:2dsolutions}(b)
clearly demonstrate the effect of the applied gate voltages on the
potential landscape of the IDQD and the device as a whole; the
result of applying a voltage on the in-plane gates is that a
significant fraction of the applied voltage is induced on the
IDQD, despite the etched trench gap. The abrupt change in the
effective permittivity from the metallic gates to the voids, from
the voids to the SiO$_2$, and from Si to SiO$_2$, causes some
definition of the gates and the IDQD in the plots. The difference
between the relative permittivity of air, Si and SiO$_2$ leads to
a potential gradient, such that the absolute value of the
potential is prone to vanish more rapidly in air, compared with Si
and SiO$_2$. However, Fig.~\ref{fig:2dsolutions}(b) clearly
demonstrates that for this particular pillar height, which matches
the device used in experiment, the potential at the IDQD is due
mainly to the electric field vectors that are on a direct path
through the trench isolation, and not the underlying substrate.
This is consistent with experimental observations and is the
preferred mechanism of device operation, since it is relatively
easier in design and theoretical analysis, compared with the case
where the majority of the electric field is through the
semiconductor base and the field lines arrive at the IDQD from
several different paths.

\begin{figure}

\epsfxsize=9.0cm

\centerline{\epsffile{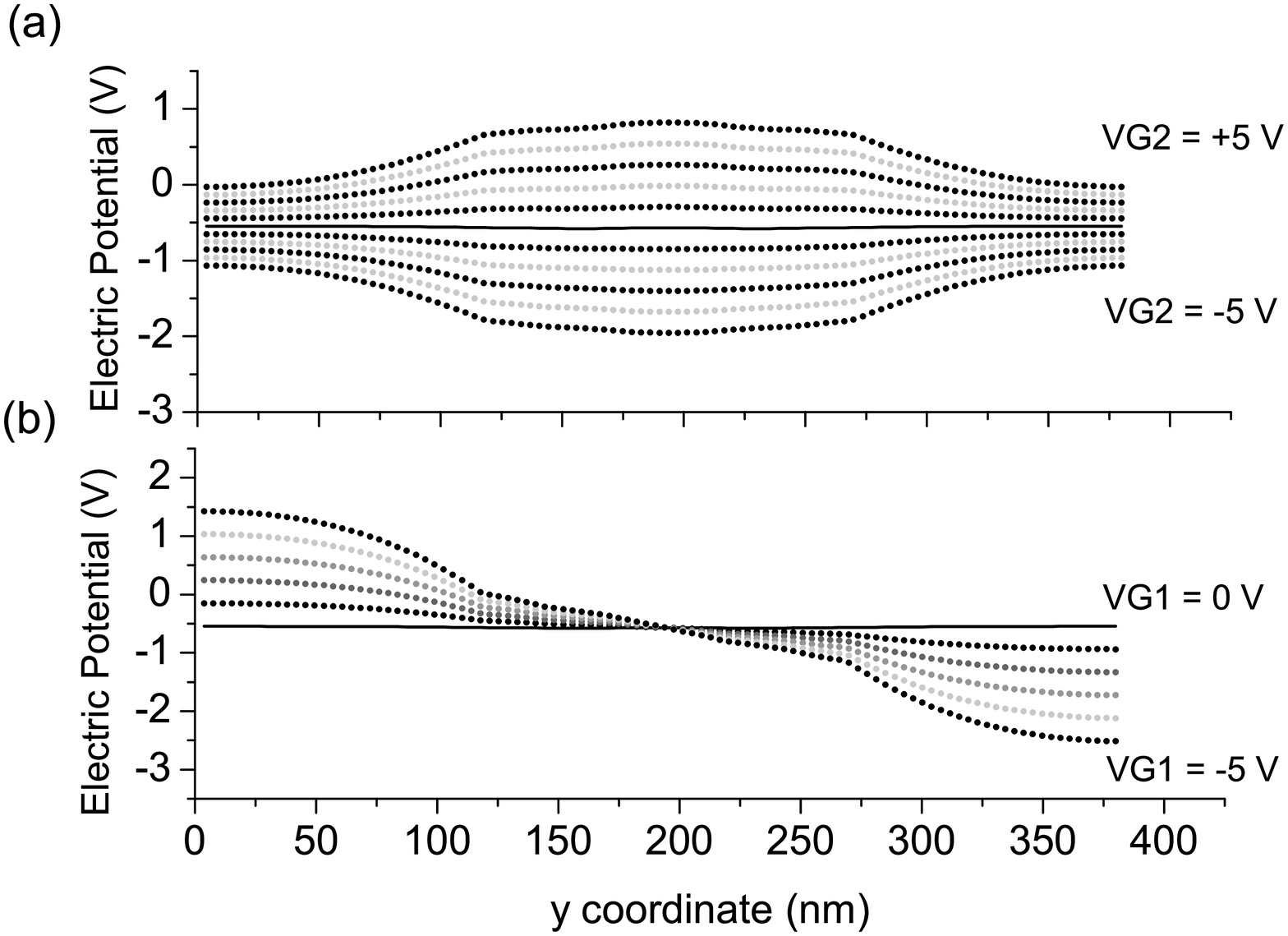}}

\caption{Cross-sectional curves through the IDQD with $x=320$~nm
and $z=420$~nm (the dots of each curve correspond to nodal points
of the computational grid). The gate G4 is set to -$4.8$~V. (a)
The electrostatic potential of the voltage on gate G2 is changed
from -$5$~V to +$5$~V. (b) The voltage applied to gate G1 is
varied from $0$~V to -$5$~V. The voltage applied to gate G3 is set
to the negative of that applied to gate G1 (in order to maximize
the electric field). } \label{fig:xscurves1}

\end{figure}

For a more quantitative analysis, we determine the electric
potential along the active region of the IDQD as a function of the
applied gate voltages. This is shown in
Figs.~\ref{fig:xscurves1}(a) and \ref{fig:xscurves1}(b), where
different gates are used to apply the in-plane electric field.
Figure~\ref{fig:xscurves1}(a) shows that the effect of varying the
voltage applied to gate G2, from -$5$~V to +$5$~V, is to induce a
voltage at the IDQD from -$2$~V to +$0.8$~V, respectively. Note
that G4 had lowered the overall potential by approximately
$0.55$~V in this case.

The data in Fig.~\ref{fig:xscurves1}(a) shows a maximum change of
$0.3$~V of the electrostatic potential at the IDQD, when the G2
gate voltage is raised or lowered by $1$~V. This field coupling
factor of $\sim$$30~\%$ is approximately one order of magnitude
greater than what was observed in experiment.\cite{Emiroglu:03}
However, the measured quantity in experiments is the SET current,
and the IDQD coupling terms are inferred from such measurements.
The task of calculating such coefficients exactly as measured in
experiment is beyond the scope of a purely electrostatic model,
since, with the SET present, the global system that must be
treated consists of interacting sub-systems of quantum
mechanically bound electrons. Therefore, we project that a
self-consistent Schr\"{o}dinger-Poisson analysis of the system
would yield results for the coupling coefficients that are closer
to actual values observed in experiment.

Our results are also consistent with the experimental
demonstrations of Ref.~\onlinecite{Emiroglu:03} where the voltage
on gate G2 is swept continuously from $-5$~V to $+5$~V, with G4
set to $\sim$$-4.8$~V, in order to demonstrate conductance
resonances of the SET currents but also resonances due to the
single electron polarization of the IDQD. (The device in
Ref.~\onlinecite{Emiroglu:03} had a slight asymmetry in the
alignment of the IDQD relative to G2, hence the ability of G2 to
polarize the IDQD.)

The abrupt changes in the potential at $115$~nm and $265$~nm are
due to the change of relative permittivity at the air-SiO$_2$
interface. The difference between the potential gradient in the
air and in the semiconductor regions is more evident in these
figures. Figure~\ref{fig:xscurves1}(b) shows the effect of
applying voltages of opposite sign to gates G1 and G3. This
results in a potential gradient across the IDQD, which has a
maximum value of $\sim$$0.007$~Vnm$^{-1}$ in these simulations.
This clearly demonstrates an effective mechanism for externally
tuning the internal potential asymmetry of the IDQD electronic
states.

In the experimental demonstrations of Ref.~\onlinecite{Gorman:05},
a voltage bias is pulsed across in-plane metallic gates, which are
placed perpendicularly to an IDQD as in our case, and this was
shown to result in the coherent oscillation of a single electron
charge present in an IDQD. This is again consistent with our
results which suggests a strong electric field is induced at the
IDQD due to the electric field at the in-plane gates.

In summary, we have successfully demonstrated, by means of
finite-element solutions to the Laplace equation, that the
electric potential and potential gradient across the confining
region of the IDQD in trench-isolated Si devices may be
manipulated effectively by the voltages applied to
capacitively-coupled in-plane gates. Our calculations show good
correlation with recent experimental demonstrations, where the
IDQD electron states are manipulated by such methods.

We thank S. Pfaendler and R. Schumann for comments and useful
discussions. SR, JG, and CB acknowledge the support of the EPSRC
through the QIP IRC. SR acknowledges the Cambridge-MIT Institute
for financial support.


\begin{references}
\bibitem{ft1} The most significant interaction is that between the SET and the IDQD
because it directly influences the read-out operation and the
degree of back-action.
\bibitem{ft2} The device elements are in fact roughly rectangular, so this
approximation introduces only a small error to the results
presented.
\bibitem{ft4} Higher-order basis functions were also implemented but had
not resulted in any significant improvement in the accuracy of the
results presented.
\end{references}
\end{document}